\documentclass[12pt]{article}
\usepackage{amssymb,amsthm,amsmath,latexsym}
\usepackage{url}
\usepackage{comment}
\usepackage{lscape}
\newtheorem{thm}{Theorem}[section]
\newtheorem{prop}[thm]{Proposition}
\newtheorem{lem}[thm]{Lemma}

\theoremstyle{remark}

\newcommand{\FF}{\mathbb{F}}
\newcommand{\ZZ}{\mathbb{Z}}

\DeclareMathOperator{\wt}{wt}

\begin{document}
\title{Ternary near-extremal self-dual codes of lengths $36$, $48$ and $60$
}

\author{
Masaaki Harada\thanks{
Research Center for Pure and Applied Mathematics,
Graduate School of Information Sciences,
Tohoku University, Sendai 980--8579, Japan.
email: \texttt{mharada@tohoku.ac.jp}.}
}

\maketitle

\begin{abstract}
For lengths $36$, $48$ and $60$, we construct new ternary near-extremal self-dual codes 
with weight enumerators for
which no ternary near-extremal self-dual codes were previously known to exist.
\end{abstract}

\section{Introduction}
Self-dual codes are one of the most interesting classes of codes.
This interest is justified by many combinatorial objects
and algebraic objects related to self-dual codes
(see e.g., \cite{SPLAG} and \cite{RS-Handbook}).

A ternary code $C$ of length $n$ is said to be \emph{self-dual} if
$C=C^\perp$, where $C^\perp$ is the dual code  of $C$.
The minimum weight $d$ of a ternary self-dual code of length $n$ is bounded by
$d \leq 3 \lfloor n/12 \rfloor +3$~\cite{MS-bound}.
A ternary  self-dual code of length $n$ and minimum weight $3 \lfloor n/12 \rfloor +3$
is called \emph{extremal}.
A ternary self-dual code of length $n$ and minimum weight
$3 \lfloor n/12 \rfloor$  is called \emph{near-extremal}.
The sparsity of ternary extremal self-dual codes of lengths divisible by $12$ 
has recently aroused interest in ternary near-extremal self-dual codes 
(see e.g., \cite{AH}, \cite{Pender}, \cite{RT}, \cite{RT2}).
In particular,  some restrictions on the weight enumerators of ternary near-extremal 
self-dual codes of length divisible by $12$ were given in~\cite{AH}.
It is a fundamental problem to determine
the weight enumerators for which there is a ternary near-extremal
self-dual code of length divisible by $12$.
In this note, we give a method for constructing ternary self-dual codes
based on negacirculant matrices.
Using the method, 
we construct new ternary near-extremal self-dual codes of lengths $36$, $48$ and $60$
with weight enumerators for
which no ternary near-extremal self-dual codes were previously known to exist.

This note is organized as follows.
In Section~\ref{Sec:2}, we give some definitions and basic results.
In Section~\ref{Sec:const}, we give a method for constructing ternary self-dual codes
of lengths divisible by $12$.
This method is based on negacirculant matrices.  
We also give some basic properties of  ternary self-dual codes constructed by this method.
In Section~\ref{Sec:36}, 
our exhaustive  search found all distinct ternary near-extremal self-dual codes of length $36$
constructed by the method given in Section~\ref{Sec:const}.
Then we found ternary near-extremal self-dual codes of length $36$ with weight enumerators for
which no ternary near-extremal self-dual codes were previously known to exist.
Similarly, in Section~\ref{Sec:48},   
we found many ternary near-extremal self-dual codes of lengths $48$ and $60$
with weight enumerators for
which no ternary near-extremal self-dual codes were previously known to exist.
Finally, in Section~\ref{Sec:72},  we give a correction of~\cite[Proposition 4.2]{AH}
which gave weight enumerators for
which ternary near-extremal self-dual codes of length $72$ were known to exist.

All computer calculations in this note were done using programs in 
\textsc{Magma}~\cite{Magma}.

\section{Preliminaries}\label{Sec:2}

In this section, we give some definitions and basic results used in this note.

Let $\FF_3=\{0,1,2\}$ denote the finite field of order $3$.
A \emph{ternary} $[n,k]$ \emph{code} $C$ is a $k$-dimensional vector subspace of $\FF_3^n$.
In this note, codes mean ternary codes.
The parameter $n$ is called the \emph{length} of $C$.
A \emph{generator matrix} of a ternary $[n,k]$ code $C$ is a $k \times n$
matrix such that the rows of the matrix generate $C$.
The \emph{weight} $\wt(x)$ of a vector $x \in \FF_3^n$ is the number of nonzero components of $x$.
A vector of a code $C$ is called a \emph{codeword} of $C$.
The minimum non-zero weight of all codewords in $C$ is called
the \emph{minimum weight} of $C$. 
The \emph{weight enumerator} of $C$ is given by $\sum_{c \in C} y^{\wt(c)}$.
Two codes $C$ and $C'$ are \emph{equivalent} if there is a
monomial matrix $P$ over $\mathbb{F}_3$ with $C' = C \cdot P$,
where $C \cdot P = \{ x P\mid  x \in C\}$.
We denote two equivalent codes $C$ and $C'$ by $C \cong C'$.

The \emph{dual} code $C^{\perp}$ of a ternary code $C$  of length $n$
is defined as:
\[
C^{\perp}=
\{x \in \FF_3^n \mid \langle x,y\rangle = 0 \text{ for all } y \in C\},
\]
where $\langle x,y\rangle = \sum_{i=1}^{n} x_i {y_i}$
for $x=(x_1,x_2,\ldots,x_n), y=(y_1,y_2,\ldots,y_n) \in \FF_3^n$.
A ternary code $C$ is said to be \emph{self-dual} if $C=C^\perp$.
Every codeword of a ternary self-dual code has weight a multiple of $3$.
A ternary self-dual code of length $n$ exists
if and only if $n \equiv 0 \pmod 4$ with $n>0$~\cite{MS-bound}.
The minimum weight $d$ of a ternary self-dual code of length $n$ is bounded by
$d \leq 3 \lfloor n/12 \rfloor +3$~\cite{MS-bound}.
A ternary  self-dual code of length $n$ and minimum weight $3 \lfloor n/12 \rfloor +3$
is called \emph{extremal}.
A ternary self-dual code of length $n$ and minimum weight
$3 \lfloor n/12 \rfloor$  is called \emph{near-extremal}.

\section{A method for constructing self-dual codes}
\label{Sec:const}

In this section, we give a method for constructing ternary self-dual codes.
We also give some basic properties of the method.

Throughout this note,
let $I_n$ denote the identity matrix of order $n$,
let $O_n$ denote the $n \times n$ zero matrix
and let
$A^T$ denote the transpose of a matrix $A$.
An $n \times n$ \emph{negacirculant} matrix has the following form:
\[
\left( \begin{array}{cccccc}
r_0&r_1&r_2& \cdots &r_{n-2} &r_{n-1}\\
2 r_{n-1}&r_0&r_1& \cdots &r_{n-3}&r_{n-2} \\
2 r_{n-2}&2 r_{n-1}&r_0& \cdots &r_{n-4}&r_{n-3} \\
\vdots &\vdots & \vdots &&\vdots& \vdots\\
2 r_1&2r_2&2 r_3& \cdots&2 r_{n-1}&r_0
\end{array}
\right).
\]
In this note, we consider ternary $[6n,3n]$ codes with the following generator matrices:
\begin{equation} \label{eq:G}
\left(
\begin{array}{ccc@{}c}
\quad & {\Large I_{3n}} & \quad &
\begin{array}{ccc}
M_{1,1} & M_{1,2} & M_{1,3} \\
M_{2,1} & M_{2,2} & M_{2,3} \\
M_{3,1} & M_{3,2} & M_{3,3} \\
\end{array}
\end{array}
\right),
\end{equation}
where $M_{i,j}$ are $n \times n$ negacirculant matrices $(i,j \in \{1,2,3\})$.
We denote the above ternary code by
\[
C(
M_{1,1}, M_{1,2}, M_{1,3},
M_{2,1}, M_{2,2}, M_{2,3}, 
M_{3,1}, M_{3,2}, M_{3,3}).
\]
Let $r_{i,j}$ denote the first row of $M_{i,j}$ $(i,j \in \{1,2,3\})$.
We denote the vector $(r_{i,1},r_{i,2},r_{i,3})$ of length $3n$ by
$r_i$ $(i\in\{1,2,3\})$.
We also denote the above ternary code by $C(r_1,r_2,r_3)$.

The following four lemmas are easily obtained by 
multiplying some rows and columns  by $2$,
and permuting some rows and columns of the generator matrix \eqref{eq:G}.

\begin{lem}\label{lem:p1}
For any vector $(a_1,a_2,a_3) \in \{1,2\}^3$,
\begin{align*}
&
C(
M_{1,1}, M_{1,2}, M_{1,3},
M_{2,1}, M_{2,2}, M_{2,3}, 
M_{3,1}, M_{3,2}, M_{3,3})
\\
&\cong
C(
a_1M_{1,1}, a_2M_{1,2}, a_3M_{1,3},
a_1M_{2,1}, a_2M_{2,2}, a_3M_{2,3}, 
a_1M_{3,1}, a_2M_{3,2}, a_3M_{3,3}).
\end{align*}
\end{lem}

\begin{lem}\label{lem:p2}
For any vector $(a_1,a_2,a_3) \in \{1,2\}^3$,
\begin{align*}
&
C(
M_{1,1}, M_{1,2}, M_{1,3},
M_{2,1}, M_{2,2}, M_{2,3}, 
M_{3,1}, M_{3,2}, M_{3,3})
\\
&\cong
C(
a_1M_{1,1}, a_1M_{1,2}, a_1M_{1,3},
a_2M_{2,1}, a_2M_{2,2}, a_2M_{2,3}, 
a_3M_{3,1}, a_3M_{3,2}, a_3M_{3,3}).
\end{align*}
\end{lem}

\begin{lem}\label{lem:p3}
For any permutation $\sigma$ of the symmetry group of degree $3$,
\begin{align*}
&
C(
M_{1,1}, M_{1,2}, M_{1,3},
M_{2,1}, M_{2,2}, M_{2,3}, 
M_{3,1}, M_{3,2}, M_{3,3})
\\
&\cong
C(
M_{1,\sigma(1)}, M_{1,\sigma(2)}, M_{1,\sigma(3)},
M_{2,\sigma(1)}, M_{2,\sigma(2)}, M_{2,\sigma(3)}, 
M_{3,\sigma(1)}, M_{3,\sigma(2)}, M_{3,\sigma(3)}).
\end{align*}
\end{lem}

\begin{lem}\label{lem:p4}
For any permutation $\sigma$ of the symmetry group of degree $3$,
\begin{align*}
&
C(
M_{1,1}, M_{1,2}, M_{1,3},
M_{2,1}, M_{2,2}, M_{2,3}, 
M_{3,1}, M_{3,2}, M_{3,3})
\\
&\cong
C(
M_{\sigma(1),1}, M_{\sigma(1),2}, M_{\sigma(1),3}, 
M_{\sigma(2),1},  M_{\sigma(2),2},  M_{\sigma(2),3},  
M_{\sigma(3),1}, M_{\sigma(3),2}, M_{\sigma(3),3}).
\end{align*}
\end{lem}


Let $C$ be a $[2n,n]$ code with generator matrix $G$.
It is trivial that $C$ is self-dual if and only if $GG^T=O_{n}$.
The following lemma is easily obtained from the above fact.

\begin{lem}\label{lem:p5}
A ternary $[6n,3n]$ code 
\[
C(
M_{1,1}, M_{1,2}, M_{1,3},
M_{2,1}, M_{2,2}, M_{2,3}, 
M_{3,1}, M_{3,2}, M_{3,3})
\]
is self-dual
if and only if
\begin{align*}
M_{i,1}M_{i,1}^T+M_{i,2}M_{i,2}^T+M_{i,3}M_{i,3}^T&=2I_{n}\ (i\in\{1,2,3\})
 \text{ and } \\
M_{i,1}M_{j,1}^T+M_{i,2}M_{j,2}^T+M_{i,3}M_{j,3}^T&=O_{n}\ (i,j\in\{1,2,3\}, i<j).
\end{align*}
\end{lem}

Since a ternary self-dual code of length $n$ exists
if and only if $n \equiv 0 \pmod 4$ with $n>0$~\cite{MS-bound},
the above method is applied to self-dual codes of length $n \equiv 0 \pmod{12}$ with $n>0$.
In order to find all  ternary self-dual codes $C(r_1,r_2,r_3)$
 of length $12m$ and minimum weight $d$
having generator matrices \eqref{eq:G}, 
it is sufficient to consider all vectors
\[
r_1=(r_{1,1},r_{1,2},r_{1,3}),
r_2=(r_{2,1},r_{2,2},r_{2,3}) \text{ and }
r_3=(r_{3,1},r_{3,2},r_{3,3})
\]
satisfying the following conditions:
\begin{itemize}

\item[(C1)]
$\wt(r_i) \equiv 2 \pmod 3$ and $\wt(r_i) \ge d-1$  $(i \in\{1,2,3\})$.

\item[(C2)]
The first nonzero component of $r_{1,j}$ is $1$ $(j \in\{1,2,3\})$ by Lemma~\ref{lem:p1}.

\item[(C3)]
The first nonzero component of $r_i$ is $1$  $(i \in\{2,3\})$ by Lemma~\ref{lem:p2}.

\item[(C4)]
For $a=(a_1,a_2,\ldots,a_{\ell}) \in \FF_3^{\ell}$, define a map $f$ from $\FF_3^{\ell}$ to 
$\ZZ$ as
$f(a)=\sum_{i=1}^{\ell} 3^{i-1} a_i$, regarding $a_i=0,1$ or $2$ as an element of $\ZZ$.
For $a,b \in \FF_3^{\ell}$,
we define an ordering $a \ge b$ based on the natural ordering of $\ZZ$
as $f(a) \ge f(b)$.
Then $f(r_{1,1}) \ge f(r_{1,2}) \ge f(r_{1,3})$ by Lemma~\ref{lem:p3}.

\item[(C5)]
$f(r_1) \ge f(r_2) \ge f(r_3)$ by Lemma~\ref{lem:p4}.

\item[(C6)]
Let $D_1$ and $D_2$ be the codes generated by the rows of the following matrices:
\[
\left(
\begin{array}{ccc}
M_{1,1} & M_{1,2} & M_{1,3} \\
\end{array}
\right)
\text{ and }
\left(
\begin{array}{ccc}
M_{1,1} & M_{1,2} & M_{1,3} \\
M_{2,1} & M_{2,2} & M_{2,3} \\
\end{array}
\right),
\]
respectively.
Then $r_2 \in D_1^\perp$ and $r_3 \in D_2^\perp$  by Lemma~\ref{lem:p5}.

\end{itemize}

To demonstrate a usefulness of this method,
we found some ternary extremal self-dual codes $C(r_1,r_2,r_3)$.
Under the conditions (C1)--(C6), our exhaustive computer search found
all distinct ternary extremal self-dual codes $C(r_1,r_2,r_3)$ 
of length $36$. 
Then we verified that every ternary extremal self-dual code $C(r_1,r_2,r_3)$ 
of length $36$ is equivalent to $C_{36}=C(r_1,r_2,r_3)$, where
\begin{align*}
r_1&=( 0, 0, 0, 0, 1, 2, 0, 1, 1, 1, 0, 2, 1, 2, 1, 0, 2, 1),\\
r_2&=( 1, 0, 1, 1, 1, 0, 1, 2, 2, 0, 2, 1, 2, 2, 1, 2, 0, 1) \text{ and }\\
r_3&=( 0, 1, 2, 1, 2, 1, 0, 2, 2, 2, 1, 2, 1, 0, 1, 2, 1, 0).
\end{align*}
For length $36$, the Pless symmetry code $P_{36}$ is a currently known ternary extremal self-dual code.
We verified that $C_{36}\cong P_{36}$.
For length $48$, 
the extended quadratic residue $QR_{48}$ and the Pless symmetry code $P_{48}$
are currently known ternary extremal self-dual codes.
We found two ternary extremal self-dual codes $C_{48}=C(r_1,r_2,r_3)$ and 
$C'_{48}=C(r'_1,r'_2,r'_3)$, where
\begin{align*}
r_1&=( 1, 1, 0, 2, 0, 1, 2, 1, 0, 1, 0, 0, 2, 2, 0, 1, 0, 1, 0, 0, 1, 1, 0, 1),\\
r_2&=( 1, 0, 2, 0, 1, 0, 0, 2, 0, 2, 0, 0, 1, 2, 2, 1, 0, 0, 2, 1, 1, 2, 2, 0 ),\\
r_3&=( 1, 0, 0, 1, 1, 0, 1, 0, 1, 2, 2, 0, 0, 0, 1, 2, 0, 0, 1, 2, 1, 2, 1, 0 ),\\
r'_1&=( 1, 0, 1, 2, 0, 0, 1, 1, 0, 1, 0, 1, 0, 0, 1, 1, 1, 2, 0, 2, 0, 2, 2, 0 ),\\
r'_2&=( 1, 0, 1, 0, 0, 1, 1, 0, 2, 1, 2, 0, 1, 2, 0, 2, 2, 1, 2, 1, 2, 1, 2, 0 ) \text{ and }\\
r'_3&=( 0, 1, 0, 1, 1, 0, 1, 2, 2, 1, 2, 1, 2, 1, 0, 2, 2, 0, 2, 2, 0, 1, 2, 0 ).
\end{align*}
We verified that $C_{48}\cong QR_{48}$ and $C'_{48}\cong P_{48}$.

\section{Near-extremal self-dual codes of length 36}
\label{Sec:36}

By the Gleason theorem (see~\cite{MS-bound}),
the possible weight enumerators of ternary near-extremal self-dual codes of length $36$
are determined as follows:
\[
\begin{split}
W_{36}=&
1
+ \alpha y^9
+( 42840 - 9 \alpha)y^{12}
+( 1400256 + 36 \alpha)y^{15}
\\&
+( 18452280- 84 \alpha)y^{18}
+( 90370368 + 126 \alpha)y^{21}
\\&
+( 162663480 - 126 \alpha)y^{24}
+( 97808480 + 84 \alpha)y^{27}
\\&
+( 16210656 - 36 \alpha)y^{30}
+( 471240 + 9 \alpha)y^{33}
+( 888 - \alpha)y^{36},
\end{split}
\]
where $\alpha$ is the number of codewords of minimum weight.
Recently,
it was shown that $\alpha=8\beta$ with $\beta \in \{1,2,\ldots,111\}$~\cite{AH}.
In addition, ternary near-extremal self-dual codes with weight enumerators
$W_{36}$ were constructed in~\cite{AH} for
\[
\alpha \in \{8 \beta \mid \beta \in \{9, 12, 14, 16, 17, \ldots, 83, 85, 90, 93\} \}.
\]
More recently,  ternary near-extremal self-dual codes with weight enumerators
$W_{36}$ were constructed in~\cite{RT} for
\[
\alpha \in \{8 \beta \mid \beta \in \{13,15\} \}.
\]

By the method given in Section~\ref{Sec:const} under the conditions (C1)--(C6),
our exhaustive  search found all distinct weight enumerators of 
ternary near-extremal self-dual codes
$C(M_{1,1}, M_{1,2}, \ldots, M_{3,3})$ of length $36$. 
The weight enumerators have the following $\alpha$ in  $W_{36}$:
\[
\alpha \in \{8 \beta \mid \beta \in \Gamma_{36} \},
\]
where
\begin{align*}
\Gamma_{36}=
\{3i,3i+1\mid i \in \{2,3,\ldots,27\}\} \cup \{85, 90, 91, 93 \}.
\end{align*}
Thus, we have the following:

\begin{prop}\label{prop:36}
There is a ternary near-extremal self-dual code with weight enumerator
$W_{36}$  for
\[
\alpha \in \{8 \beta \mid \beta \in \{6, 7, 10, 91\} \}.
\]
\end{prop}

As an example, we give in Table~\ref{Tab:36} $r_1,r_2,r_3$ in
ternary near-extremal self-dual codes 
$C(r_1,r_2,r_3)=C_1,C_2,C_3$ and $C_4$ with weight enumerators
$W_{36}$  for $\alpha=8\beta$, where $\beta=6, 7, 10$ and $91$, respectively.

\begin{table}[thb]
\caption{Ternary near-extremal self-dual codes of length $36$}
\label{Tab:36}
\centering
\medskip
{\small
\begin{tabular}{c|c|l}
\noalign{\hrule height1pt}
Code & $\beta$&  \multicolumn{1}{c}{$r_1,r_2,r_3$} \\
\hline
$C_1$ 
&6&$r_1=(0,0,1,0,0,2,0,0,1,1,1,1,1,2,2,0,1,1)$\\
&&$r_2=(0,1,2,2,2,1,2,1,1,0,2,2,1,0,2,2,0,1)$\\
&&$r_3=(1,2,0,0,2,1,2,0,1,1,1,2,2,1,1,1,2,0)$\\
\hline
$C_2$
&7&$r_1=(0,0,0,0,1,1,0,0,1,1,0,1,0,1,1,0,0,1)$\\
&&$r_2=(1,1,2,1,2,0,2,0,2,1,0,1,1,2,1,2,1,0)$\\
&&$r_3=(1,0,1,0,2,1,1,2,1,0,0,2,0,1,2,2,0,0)$ \\
\hline
$C_3$ 
&10&$r_1=(0,0,0,1,1,1,0,0,1,2,2,0,1,2,2,1,2,0)$\\
&&$r_2=(1,0,2,2,2,2,1,2,0,0,1,0,0,0,2,1,1,0)$\\
&&$r_3=(1,1,2,2,1,1,1,1,1,1,0,2,1,1,1,0,0,0)$\\
\hline
$C_4$ 
&91&$r_1=(0,0,1,0,1,1,0,0,1,2,0,1,0,1,0,0,2,0)$\\
&&$r_2=(0,1,2,2,2,2,1,1,0,1,0,1,2,1,2,2,1,0)$\\
&&$r_3=(1,1,2,2,1,1,1,1,2,0,0,2,2,1,2,2,0,0)$\\
\noalign{\hrule height1pt}
\end{tabular}
}
\end{table}

In order to construct more new ternary near-extremal self-dual codes, we
consider the following method.
Let $C$ be a ternary self-dual code of length $n$.  
Let $x$ be a vector of length $n$ such that $\wt(x) \equiv 0 \pmod 3$ and $x \not\in C$.
Then it is trivial that
\[
N(C,x)=
\langle (C \cap \langle x \rangle^\perp), x \rangle
\]
is a self-dual code.
Using this method, we found 
ternary near-extremal self-dual codes 
$C_5=N(C_2,x_1)$ and $C_6=N(C_2,x_2)$ 
with weight enumerators
$W_{36}$  for $\alpha=8\beta$, where $\beta=8$ and $11$, 
where
\begin{align*}
x_1&=(0,0,\ldots,0,1, 2, 0, 1, 2, 2, 0, 2, 2, 0, 2, 1, 2, 0, 1, 0, 2, 0) \text{ and }\\
x_2&=(0,0,\ldots,0,1, 1, 1, 1, 1, 0, 0, 2, 0, 0, 1, 1, 2, 0, 2, 2, 0, 2),
\end{align*}
respectively.

Combined with Proposition~\ref{prop:36}, we have the following:

\begin{prop}
There is a ternary near-extremal self-dual code with weight enumerator
$W_{36}$  for
\[
\alpha \in \{8 \beta \mid \beta \in \{6, 7, 8, 10, 11, 91\} \}.
\]
\end{prop}

Therefore, it remains to determine whether there is a ternary near-extremal self-dual code 
with weight enumerator
$W_{36}$  for
\[
\alpha \in \{8 \beta \mid \beta \in \{1,2,3,4,5,84,86,87,88,89,92,94,95,\ldots,111\} \}.
\]

\section{Near-extremal self-dual codes of lengths 48 and 60}
\label{Sec:48}

\subsection{Near-extremal self-dual codes of length 48}

By the Gleason theorem (see~\cite{MS-bound}),
the possible weight enumerators of ternary near-extremal self-dual codes of length $48$
are  determined as follows:
\[
\begin{split}
W_{48}=&
1
+\alpha                  y^{12}
+( 415104 - 12\alpha       )y^{15}
+( 20167136 + 66\alpha     )y^{18}
\\&
+( 497709696 - 220\alpha   )y^{21}
+( 5745355200 + 495\alpha  )y^{24}
\\&
+( 31815369344 - 792\alpha )y^{27}
+( 83368657152 + 924\alpha )y^{30}
\\&
+( 99755406432 - 792\alpha )y^{33}
+( 50852523072 + 495\alpha )y^{36}
\\&
+( 9794378880 - 220\alpha  )y^{39}
+( 573051072 + 66\alpha    )y^{42}
\\&
+( 6503296 - 12\alpha      )y^{45}
+( 96 + \alpha             )y^{48},
\end{split}
\]
where $\alpha$ is the number of codewords of minimum weight.
Recently,
it was shown that $\alpha=8\beta$ with $\beta \in \{1,2,\ldots,4324\}$~\cite{AH}.
In addition, ternary near-extremal self-dual codes with weight enumerators
$W_{48}$ were constructed in~\cite{AH} for
\[
\alpha \in 
\left\{48 \beta \mid \beta \in \Gamma_{48,1}\right\} \cup
\left\{8 \beta \mid \beta \in \Gamma_{48,2}\right\},
\]
where
\begin{align*}
\Gamma_{48,1}&=
\left\{\begin{array}{l}
33, 34, 36, 37,\ldots, 107, 110, 113, 115, 116, \\
117, 118, 123, 126, 132, 142, 166, 246 
\end{array}\right\} \text{ and }
\\
\Gamma_{48,2}&=
\left\{\begin{array}{l}
180, 181, 182, 184, 188, 210, 212, 215, \\
217, 218,\ldots , 275, 277, 278, 281
\end{array}\right\}
\\&\quad
\setminus \{222,228,234,240,246,252,258,264,270,276\}.
\end{align*}
More recently,  ternary near-extremal self-dual codes with weight enumerators
$W_{48}$ were constructed in~\cite{RT2} for
\[
\alpha \in 
\left\{8 \beta \mid \beta \in \Gamma_{48,3}\right\},
\]
where $\Gamma_{48,3}$ is listed in Table~\ref{Tab:483}.
In addition, ternary near-extremal self-dual codes with weight enumerators
$W_{48}$ were constructed in~\cite{Pender} for
\[
\alpha \in 
\left\{8 \beta \mid \beta \in \Gamma_{48,4}\right\},
\]
where $\Gamma_{48,4}$ is listed in Table~\ref{Tab:484}.

\begin{table}[thb]
\caption{$\Gamma_{48,3}$}
\label{Tab:483}
\centering
\medskip
{\small
\begin{tabular}{l}
\noalign{\hrule height1pt}
$313, 320, 323, 324, 326, 329, 331, 332, 333, 334,
337,\ldots,341,$\\
$343,\ldots,468,
470,\ldots,480,
482,\ldots,486,
488,\ldots,494,496,\ldots,500$,\\
$503, 504, 506, 512, 516, 518, 522, 524, 528, 536, 554, 560$\\
\noalign{\hrule height1pt}
\end{tabular}
}
\end{table}

\begin{table}[thb]
\caption{$\Gamma_{48,4}$}
\label{Tab:484}
\centering
\medskip
{\small
\begin{tabular}{l}
\noalign{\hrule height1pt}
$279, 285, 291, 297, 298, 303, 308, 309, 315, 321, 327, 328, 469, 481,$\\
$487, 495, 501, 507, 508, 509, 513, 514, 517, 519, 520, 521, 525, 526,$\\
$531, 532, 537, 543, 548, 549, 555, 557, 561, 565, 567, 568, 572, 573,$\\
$579, 580, 581, 585, 597, 598, 604, 609, 615, 620, 621, 633, 639, 644,$\\
$645, 648, 652, 657, 666, 669, 672, 681, 693, 700, 711, 714, 717, 720,$\\
$729, 732, 752, 753, 777, 780, 789, 804, 873, 900, 924, 933, 1092$\\
\noalign{\hrule height1pt}
\end{tabular}
}
\end{table}

Under the conditions (C1)--(C6),  our non-exhaustive search found
$20$ ternary near-extremal self-dual codes $C(r_1,r_2,r_3)$ of length $48$
with weight enumerators for
which no ternary near-extremal self-dual codes were previously known to exist
as follows.

\begin{prop}
There is a ternary near-extremal self-dual code with weight enumerator
$W_{48}$  for
\[
\alpha \in \{8 \beta \mid \beta \in \Gamma_{48,5} \},
\]
where 
\[
\Gamma_{48,5}=
\left\{\begin{array}{l}
280, 284, 292, 296, 304, 316, 544, 556, 584, 592, \\
596, 616, 628, 640, 668, 676, 692, 936, 1044, 1300\\
\end{array}\right\}.
\]
\end{prop}

\begin{table}[thbp]
\caption{Ternary near-extremal self-dual codes of length $48$}
\label{Tab:48}
\centering
\medskip
{\footnotesize
\begin{tabular}{c|l}
\noalign{\hrule height1pt}
$\beta$&  \multicolumn{1}{c}{$r_1,r_2,r_3$} \\
\hline
280
&$r_1=(0,0,0,1,0,2,1,0,1,0,2,0,0,2,0,0,1,0,2,2,2,1,0,0)$\\
&$r_2=(0,0,1,0,2,1,0,0,0,1,1,2,1,0,1,1,1,2,2,0,1,1,0,0)$\\
&$r_3=(1,0,2,2,1,2,0,0,0,1,2,1,2,0,1,0,2,2,0,1,0,1,0,0)$\\
\hline
284
&$r_1=(0,0,0,1,0,0,1,0,0,1,2,0,2,2,0,0,1,2,1,0,1,2,0,0)$\\
&$r_2=(1,0,1,0,0,1,0,1,2,0,1,0,1,2,1,2,1,0,2,1,0,2,0,0)$\\
&$r_3=(1,2,1,2,0,0,0,2,0,0,2,0,2,1,2,0,2,2,0,2,1,1,0,0)$\\
\hline
292
&$r_1=(0,0,0,1,0,0,1,0,0,0,1,2,0,2,0,0,1,2,1,1,1,1,0,0)$\\
&$r_2=(1,0,2,1,2,0,1,0,1,1,2,0,1,1,0,1,0,0,2,2,2,0,0,0)$\\
&$r_3=(1,0,1,2,1,1,2,1,2,1,0,2,1,0,1,0,2,2,1,1,2,0,0,0)$\\
\hline
296
&$r_1=(0,0,0,1,0,1,1,0,1,1,0,2,1,0,1,0,1,2,0,0,0,2,0,0)$\\
&$r_2=(1,1,2,0,2,1,2,2,0,0,1,2,0,2,0,2,0,0,2,2,0,1,0,0)$\\
&$r_3=(1,2,0,2,0,1,0,0,2,2,2,1,0,2,0,0,2,2,2,1,2,0,0,0)$\\
\hline
304
&$r_1=(0,0,0,1,0,1,1,0,1,2,0,0,0,1,1,0,0,1,0,2,1,0,1,0)$\\
&$r_2=(1,2,0,1,0,0,2,1,0,1,2,0,1,0,0,2,1,2,2,2,0,0,1,0)$\\
&$r_3=(1,1,1,2,0,1,1,0,0,2,1,1,2,2,2,0,1,2,1,2,0,1,0,0)$\\
\hline
316
&$r_1=(0,0,0,1,0,2,0,0,1,1,0,2,2,1,0,0,1,0,1,0,2,1,0,0)$\\
&$r_2=(1,0,0,2,1,2,0,2,1,1,2,0,1,1,0,0,2,2,0,0,2,1,0,0)$\\
&$r_3=(1,2,2,1,1,2,0,1,0,2,0,1,0,0,2,2,1,1,0,0,1,0,0,0)$\\
\hline
544
&$r_1=(1,1,0,0,2,2,1,0,0,0,0,1,1,2,1,0,0,0,0,1,0,2,0,0)$\\
&$r_2=(0,1,0,2,0,0,0,0,2,1,2,0,1,2,1,1,2,1,1,0,2,1,0,0)$\\
&$r_3=(1,2,2,2,0,1,1,1,0,0,1,0,2,0,2,0,1,2,2,2,0,0,0,0)$\\
\hline
556
&$r_1=(1,2,0,0,0,0,0,1,1,2,0,0,0,0,0,1,1,2,0,1,2,0,2,0)$\\
&$r_2=(1,1,2,2,0,2,1,2,0,2,0,1,1,1,0,1,1,0,2,1,2,0,1,0)$\\
&$r_3=(1,1,1,0,2,2,1,1,2,2,2,2,0,0,2,2,1,2,0,2,1,0,0,0)$\\
\hline
584
&$r_1=(0,0,0,1,1,2,1,0,0,0,0,1,1,2,1,0,1,0,0,1,2,0,0,0)$\\
&$r_2=(0,1,0,1,1,1,2,0,0,2,2,0,2,2,2,0,2,1,2,2,0,0,0,0)$\\
&$r_3=(0,1,1,0,1,1,1,0,0,2,0,2,2,2,1,0,1,2,1,1,0,0,0,0)$\\
\hline
592
&$r_1=(1,0,0,0,2,1,2,0,0,0,1,1,0,2,0,0,0,1,2,1,0,1,0,0)$\\
&$r_2=(1,1,2,2,0,0,2,2,2,0,2,1,2,2,2,2,1,0,2,2,1,0,0,0)$\\
&$r_3=(1,1,2,1,2,1,2,0,1,2,2,2,1,1,0,2,1,2,0,0,1,0,0,0)$\\
\hline
596
&$r_1=(1,0,2,2,0,1,0,1,0,0,1,2,2,2,2,0,1,0,1,1,0,0,1,0)$\\
&$r_2=(1,1,1,1,1,1,1,0,2,0,1,2,1,2,2,0,1,2,0,2,0,1,0,0)$\\
&$r_3=(1,0,1,1,0,1,0,1,0,2,2,2,1,0,1,2,2,0,2,0,1,0,0,0)$\\
\hline
616
&$r_1=(0,0,1,1,2,2,0,1,1,2,2,0,1,1,0,1,1,2,1,2,0,1,0,1)$\\
&$r_2=(0,1,2,1,1,0,1,0,0,2,2,2,2,2,1,1,0,0,0,1,0,0,0,1)$\\
&$r_3=(1,0,0,1,2,2,0,0,2,0,1,2,2,2,1,1,2,2,2,2,2,0,1,0)$\\
\hline
628
&$r_1=(1,2,2,1,2,1,2,0,1,2,1,2,0,1,2,0,1,1,2,2,2,1,1,0)$\\
&$r_2=(1,2,2,1,1,0,0,0,1,1,0,0,0,2,1,1,0,0,0,0,1,0,0,0)$\\
&$r_3=(1,1,2,0,1,1,0,0,0,0,0,1,1,0,1,0,2,1,2,0,0,0,0,0)$\\
\hline
640
&$r_1=(0,1,2,0,1,0,1,1,1,1,1,0,1,2,0,1,0,1,0,0,0,2,0,1)$\\
&$r_2=(1,1,1,0,0,1,1,1,0,0,0,1,0,0,0,1,1,1,0,2,2,1,2,0)$\\
&$r_3=(0,1,0,2,0,1,0,0,1,0,0,2,1,2,2,1,2,1,0,2,2,0,1,0)$\\
\noalign{\hrule height1pt}
\end{tabular}
}
\end{table}

\begin{table}[thbp]
\caption{Ternary near-extremal self-dual codes of length $48$}
\label{Tab:48-2}
\centering
\medskip
{\footnotesize
\begin{tabular}{c|l}
\noalign{\hrule height1pt}
$\beta$&  \multicolumn{1}{c}{$r_1,r_2,r_3$} \\
\hline
668
&$r_1=(1,2,1,2,1,0,2,1,1,2,1,1,0,1,1,1,0,1,1,1,1,2,2,0)$\\
&$r_2=(1,0,1,1,2,0,1,2,0,0,1,0,0,1,0,2,0,0,2,0,0,0,1,0)$\\
&$r_3=(1,2,1,0,2,1,2,1,1,2,1,2,2,2,0,2,1,1,2,2,2,2,0,0)$\\
\hline
676
&$r_1=(0,1,2,2,0,1,0,1,0,1,2,1,0,1,0,1,1,0,0,2,1,0,2,0)$\\
&$r_2=(1,0,1,1,2,0,2,1,1,2,1,0,2,1,1,0,0,0,0,0,0,1,1,0)$\\
&$r_3=(1,2,0,0,0,1,0,0,1,0,0,1,0,0,0,2,0,1,1,2,2,1,0,0)$\\
\hline
692
&$r_1=(1,0,0,1,0,2,1,0,0,1,0,2,2,0,1,0,1,0,0,0,2,0,1,0)$\\
&$r_2=(1,0,0,1,1,0,2,0,0,0,0,2,0,2,0,2,1,0,2,1,0,2,0,0)$\\
&$r_3=(0,1,0,0,0,2,0,2,0,1,2,0,1,0,1,0,1,0,2,2,0,1,0,0)$\\
\hline
936
&$r_1=(1,2,0,2,2,1,1,0,1,0,2,1,1,1,1,0,1,1,2,0,2,2,0,0)$\\
&$r_2=(1,1,2,0,2,2,1,0,1,2,2,2,0,1,2,1,1,0,2,1,0,1,0,0)$\\
&$r_3=(1,2,1,2,1,2,1,2,0,1,1,2,1,2,2,2,2,2,1,1,2,0,0,0)$\\
\hline
1044
&$r_1=(1,1,1,0,1,0,1,1,0,0,1,2,2,1,0,1,1,0,0,0,1,0,0,1)$\\
&$r_2=(1,0,2,1,1,0,0,1,1,1,1,0,1,0,1,1,1,0,1,1,2,2,2,0)$\\
&$r_3=(1,1,2,2,2,0,2,0,0,0,1,0,0,1,2,0,2,1,1,2,0,2,0,0)$\\
\hline
1300
&$r_1=(0,1,1,0,1,2,2,1,1,1,2,2,0,1,0,1,0,0,1,2,2,2,1,0)$\\
&$r_2=(1,0,0,0,2,1,1,1,2,2,1,0,0,0,2,0,0,1,1,2,2,1,0,0)$\\
&$r_3=(0,1,0,1,2,2,1,1,1,1,2,0,2,2,1,1,1,0,2,2,1,0,0,0)$\\
\noalign{\hrule height1pt}
\end{tabular}
}
\end{table}



In Tables~\ref{Tab:48} and~\ref{Tab:48-2}, we give  $r_1,r_2,r_3$ in
the $20$ ternary near-extremal self-dual codes $C(r_1,r_2,r_3)$ of length $48$.

\subsection{Near-extremal self-dual codes of length 60}

By the Gleason theorem (see~\cite{MS-bound}),
the possible weight enumerators of ternary near-extremal self-dual codes of length $60$
are  determined as follows:
\[
\begin{split}
W_{60}=&1
+\alpha  y^{15}
+(3901080 - 15\alpha ) y^{18}
+(241456320 + 105\alpha)  y^{21}
\\&
+(8824242960 - 455\alpha)  y^{24}
+(172074038080 + 1365\alpha)  y^{27}\\&
+(1850359081824 - 3003\alpha)  y^{30}
+(11014750094040 + 5005\alpha) y^{33}\\&
+ \cdots
+(71451360 + 15\alpha) y^{57}+(41184 - \alpha )y^{60},
\end{split}
\]
where $\alpha$ is the number of codewords of minimum weight.
Recently,
it was shown that  $\alpha=8\beta$ with $\beta \in \{1,2,\ldots,5148\}$~\cite{AH}.
In addition, ternary near-extremal self-dual codes with weight enumerators
$W_{60}$ were constructed in~\cite{AH} for
\[
\alpha \in 
\{24\beta \mid \beta \in \Gamma_{60,1}\} \cup \{8\beta \mid \beta \in \Gamma_{60,2}\}, 
\]
where $\Gamma_{60,1}$ and $\Gamma_{60,2}$ are listed in~\cite[Tables~22 and 27]{AH}, respectively.

Under the conditions (C1)--(C6),  our non-exhaustive search found
$275$ ternary near-extremal self-dual codes $C(r_1,r_2,r_3)$ of length $60$
with weight enumerators for
which no ternary near-extremal self-dual codes were previously known to exist
as follows.

\begin{prop}
There is a ternary near-extremal self-dual code with weight enumerator
$W_{60}$  for
\[
\alpha \in \{8 \beta \mid \beta \in \Gamma_{60,3} \},
\]
where $\Gamma_{60,3}$ is listed in Table~\ref{Tab:60}.
\end{prop}

\begin{table}[thb]
\caption{$\Gamma_{60,3}$}
\label{Tab:60}
\centering
\medskip
{\small
\begin{tabular}{l}
\noalign{\hrule height1pt}
$2766, 2770, 2780, 2781, 2785, 2786, 2788, 2791, 2795, 2798, 2800, 2801, 2803,$\\
$2806, 2810, 2811, 2815, 2816, 2818, 2821, 2825, 2826, 2828, 2830, 2831, 2833,$\\
$2836, 2840, 2841, 2843, 2845, 2846, 2848, 2851, 2855, 2856, 2858, 2860, 2861,$\\
$2863, 2866, 2870, 2871, 2873, 2875, 2876, 2878, 2881, 2885, 2886, 2888, 2890,$\\
$2891, 2893, 2896, 2900, 2901, 2903, 2905, 2906, 2908, 2911, 2915, 2916, 2918,$\\
$2920, 2921, 2923, 2926, 2930, 2931, 2933, 2935, 2936, 2938, 2941, 2945, 2946,$\\
$2948, 2950, 2951, 2953, 2956, 2960, 2961, 2963, 2965, 2966, 2968, 2971, 2975,$\\
$2976, 2978, 2980, 2981, 2983, 2986, 2990, 2991, 2993, 2995, 2996, 2998, 3001,$\\
$3005, 3006, 3008, 3010, 3011, 3013, 3016, 3020, 3021, 3023, 3025, 3026, 3028,$\\
$3031, 3035, 3036, 3038, 3040, 3041, 3043, 3046, 3050, 3051, 3053, 3055, 3056,$\\
$3058, 3061, 3065, 3066, 3068, 3070, 3071, 3073, 3076, 3080, 3081, 3083, 3085,$\\
$3086, 3088, 3091, 3095, 3096, 3098, 3100, 3101, 3103, 3106, 3110, 3111, 3113,$\\
$3115, 3116, 3118, 3121, 3125, 3126, 3128, 3130, 3131, 3133, 3136, 3140, 3141,$\\
$3143, 3145, 3146, 3148, 3151, 3155, 3156, 3158, 3160, 3161, 3163, 3166, 3170,$\\
$3171, 3173, 3175, 3176, 3178, 3181, 3185, 3186, 3188, 3190, 3191, 3193, 3196,$\\
$3200, 3201, 3203, 3205, 3206, 3208, 3211, 3215, 3216, 3218, 3220, 3221, 3223,$\\
$3226, 3230, 3231, 3233, 3235, 3236, 3238, 3241, 3245, 3246, 3248, 3250, 3251,$\\
$3253, 3256, 3260, 3261, 3263, 3265, 3266, 3268, 3271, 3275, 3276, 3278, 3280,$\\
$3281, 3283, 3286, 3290, 3291, 3293, 3295, 3296, 3298, 3301, 3305, 3306, 3308,$\\
$3310, 3311, 3313, 3316, 3320, 3321, 3323, 3325, 3326, 3328, 3331, 3335, 3336,$\\
$3338, 3340, 3341, 3343, 3346, 3351, 3353, 3356, 3361, 3365, 3366, 3371, 3373,$\\
$3376, 3385$ \\
\noalign{\hrule height1pt}
\end{tabular}
}
\end{table}

For the $275$ ternary near-extremal self-dual codes $C(r_1,r_2,r_3)$ of length $60$,
$(r_1,r_2,r_3)$ are available at
\url{https://www.math.is.tohoku.ac.jp/~mharada/F3-9nega}.

\section{Correction of~\cite[Proposition 4.2]{AH}}
\label{Sec:72}

In this section, we give a correction of~\cite[Proposition 4.2]{AH}.

By the Gleason theorem (see~\cite{MS-bound}),
the possible weight enumerators of ternary near-extremal self-dual codes of length $72$
are  determined as follows:
\[
\begin{split}
W_{72}=&1
+\alpha  y^{18}
+(36213408  - 18 \alpha ) y^{21}
+(2634060240 + 153 \alpha ) y^{24}\\&
+(126284566912 - 816 \alpha ) y^{27}
+(3525613242624  + 3060 \alpha ) y^{30}\\&
+(59358705673680  - 8568 \alpha ) y^{33}
+(607797076070496  + 18564 \alpha ) y^{36}\\&
+\cdots 
+(707807520 - 18 \alpha ) y^{69}
+(- 115728 + \alpha ) y^{72},
\end{split}
\]
%
%
where $\alpha$ is the number of codewords of minimum weight.
Recently,
it was shown that  $\alpha=8\beta$ with $\beta \in \{14466,14467,\ldots,251482\}$~\cite{AH}.

In~\cite[Proposition 4.2]{AH}, the existence of 
ternary near-extremal self-dual codes with weight enumerators
$W_{72}$ for
\begin{equation}\label{eq:WE72}
\alpha \in \{24\beta \mid \beta \in  \{8914,14910\} \cup \Gamma_{72} \cup \Gamma'_{72}
\} \cup \{8\beta \mid \beta \in \Delta_{72}\}
\end{equation}
was claimed (see~\cite{AH} for the sets $\Gamma_{72}, \Gamma'_{72},\Delta_{72}$).
However, it is clear from~\cite[(10)]{AH} that \eqref{eq:WE72} is incorrect and the correct result is
\[
\alpha \in \Gamma_{72} \cup
\{24\beta \mid \beta \in  \{8914,14910\} \cup  \Gamma'_{72}
\} \cup \{8\beta \mid \beta \in \Delta_{72}\}.
\]

Similar to lengths $36,48$ and $60$, 
by the method given in Section~\ref{Sec:const} under the conditions (C1)--(C6),
we tried to find a ternary near-extremal self-dual code of length $72$.
However, an extensive search failed to discover such a code.

\bigskip
\noindent
\textbf{Acknowledgments.}
This work was supported by JSPS KAKENHI Grant Number 23K25784.

\end{document}